\documentclass[aip,rsi,amsmath,amssymb,reprint]{revtex4-1}
\usepackage{graphicx}
\usepackage{dcolumn}
\usepackage{bm}

\begin{document}

\preprint{AIP/123-QED}

\title{Four-wave mixing in a silicon microring resonator using a self-pumping geometry.}

\author{Micol Previde Massara}
\affiliation{Dipartimento di Fisica, Universit\'a degli Studi di  Pavia, via A. Bassi 6, 27100 Pavia, Italy}

\author{Federico Andrea Sabattoli}
\affiliation{Dipartimento di Fisica, Universit\'a degli Studi di  Pavia, via A. Bassi 6, 27100 Pavia, Italy}

\author{Federico Pirzio}
\affiliation{Dipartimento di Ingegneria Industriale e dell'Informazione, Universit\'a degli Studi di Pavia, via Ferrata 1, 27100 Pavia, Italy}

\author{Matteo Galli}
\affiliation{Dipartimento di Fisica, Universit\'a degli Studi di  Pavia, via A. Bassi 6, 27100 Pavia, Italy}

\author{Daniele Bajoni}
\affiliation{Dipartimento di Ingegneria Industriale e dell'Informazione, Universit\'a degli Studi di Pavia, via Ferrata 1, 27100 Pavia, Italy}

          
\begin{abstract}
We report on four-wave mixing in a silicon microring resonator using a self-pumping scheme instead of an external laser. The ring resonator is inserted in an external-loop cavity with a fibered semiconductor amplifier as a source of gain. The silicon microring acts as a filter and we observe lasing in one of the microring's resonances. We study correlations between signal and idler generated beams using a Joint Spectral Density experiment.
\end{abstract}

\maketitle

Silicon ring resonators are microscopic devices integrated on a silicon chip that have been shown to be efficient for on-chip optical nonlinearities. Low-power optical nonlinearities in silicon integrated devices are used to achieve a variety of effects including all-optical switching \cite{Almeida2004}, optical bistability \cite{AlmeidaLipsonOL2004} and four-wave mixing (FWM) \cite{Turner2008}. FWM in particular has been shown to be greatly enhanced by the light confinement in microring resonators, to the point of being able to produce photon pairs with MHz rates \cite{Azzini2012OE,Clemmen2009} and act as microscopic, integrated sources of entangled photons \cite{Grassani:Optica:2015,Wakabayashi:2015:ringtimebin,Harris:PRX:2014}. As a result, ring resonators are among the most promising sources of nonclassical states of light for use in quantum technologies \cite{Caspani_review}.

An important hurdle that still remains to be solved to facilitate the widespread adoption of microring resonators as quantum optical sources and for the on-chip generation of frequency combs \cite{Pasquazi:2013,Reimer2014} is the need for an external optical pump. Silicon microrings are resonant structures, therefore have discrete sharp spectral resonances. The pump needs to be tunable and, sometimes, even actively tuned to accommodate spectral changes in the ring’s resonances due to thermal or power fluctuations, that could be hundreds of times the resonance linewidth \cite{Carmon:2004,Peccianti:2012}. In silicon resonators tuning is particularly relevant in the case of high-Q resonators \cite{Jiang:Painter:2015:microdiskcoinc} and high pumping powers \cite{Azzini2012OL}.
The need of a tunable source is a cumbersome and expensive requirement for devices that should be economic and nimble to operate. In this work we relax this requirement by demonstrating that FWM can be achieved in a silicon microring resonator using self-pumping \cite{Pasquazi:2013,Reimer:Morandotti:2015:NatCommselfpumping}. We insert the resonator inside an external fiber-loop cavity including an amplifier: in this geometry gain is spectrally filtered by the microring resonances and we show that lasing can be achieved with enough power to measure FWM emission.



\begin{figure}
\includegraphics[width=\columnwidth]{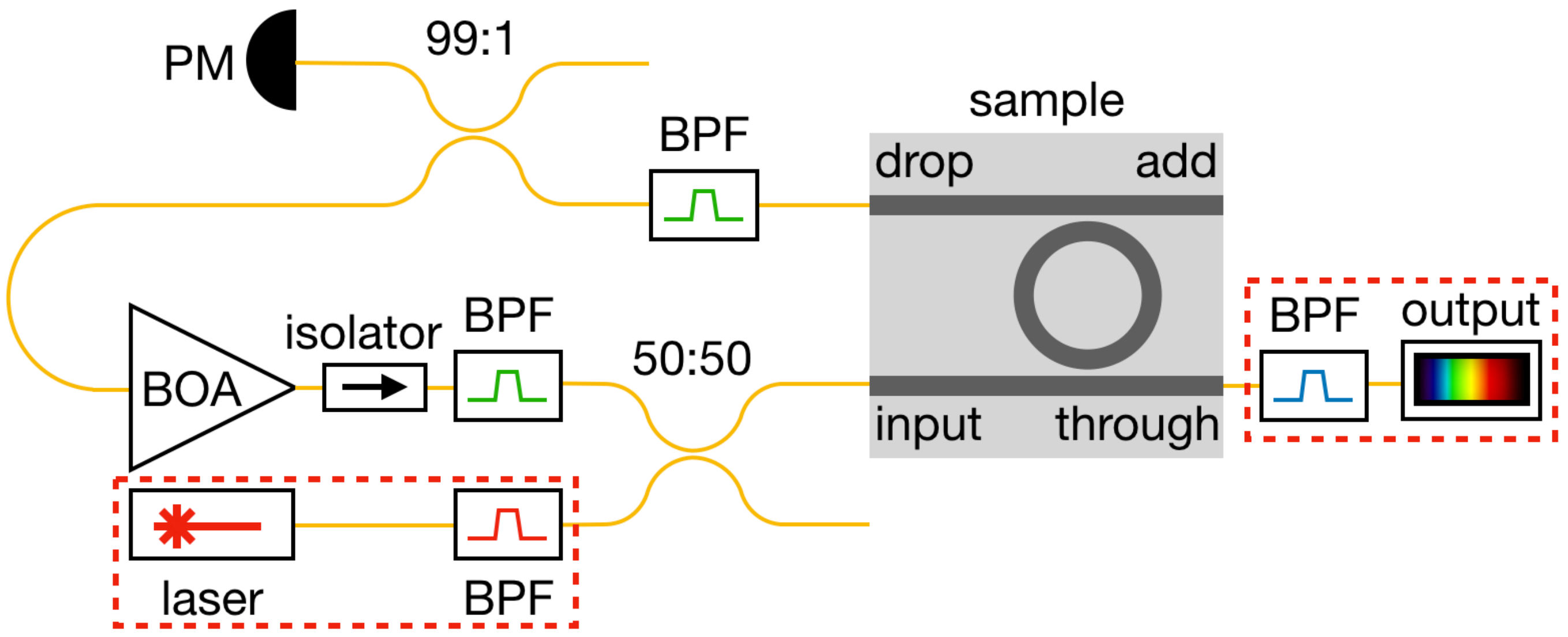}
\caption{Schematic view of the built cavity. BPF stands for band-pass filter, PM for power meter and BOA for booster optical amplifier. A schematic view of the add-drop ring resonator used is shown as an inset. The elements inside the dashed red rectangles are connected to the cavity for the stimulated FWM experiment only. The scheme is not in scale.}
\label{fig:1}
\end{figure}

The scheme of the experimental apparatus used for the cavity is shown in in Fig. \ref{fig:1}. The components inside the dashed red rectangles are outside the cavity and connected to it for the stimulated FWM experiment only.
The sample was fabricated at the INtegrated Photonic TEchnologies Center (INPHOTEC) in Pisa and realized by a e-beam lithography process on a 6-inch silicon-on-insulator (SOI) wafer. A buried oxide layer is covered by a 220-nm-thick silicon layer, with a bulk refractive index $n_{Si}$ = 3.48 at 1550 nm. The SOI structure is then coated by an oxide cladding (bulk index $n_{SiO_{2}}$ = 1.46). The ridge waveguide cross-section area is 220 x 480 nm\textsuperscript{2} and is designed to support a single guided mode.

\begin{figure}
\includegraphics[width=\columnwidth]{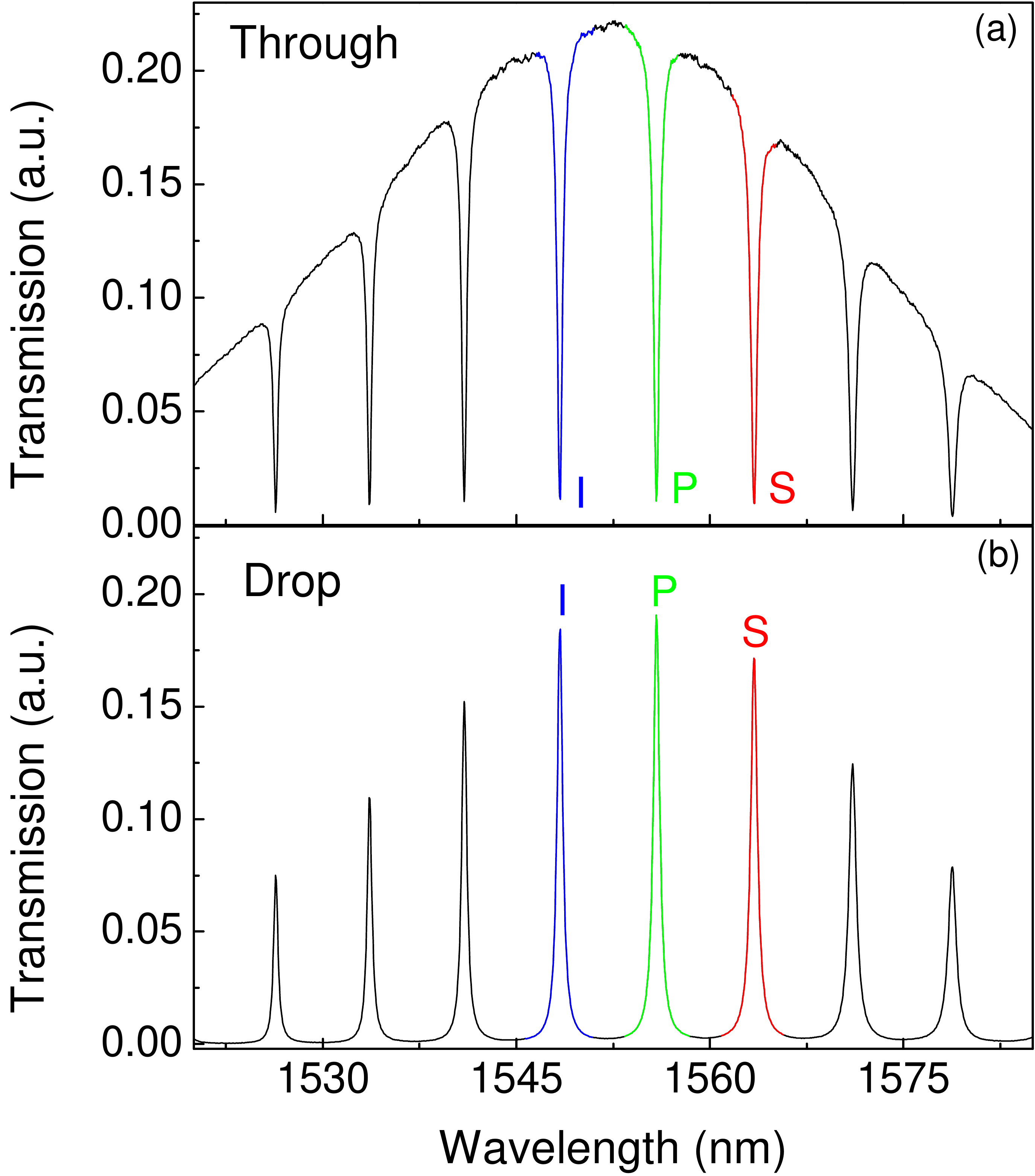}
\caption{Transmission spectra of the (a) through and (b) drop ports of the add-drop ring resonator used in the cavity. The resolution of the spectra is 50 pm. From the through-port spectrum, it can be seen that the ring resonances go down to below 5\% of transmission. This means that the add-drop ring resonator is close to the critical coupling condition. I, P and S stand, respectively, for the idler, pump and signal resonances used in the stimulated FWM experiment. }
\label{fig:2}
\end{figure}

Given the persisting difficulty of achieving optical gain in silicon, the idea is to take advantage of an external source of gain and build a closed-loop cavity with the source of entangled photons inside the loop. The building block is made out of a ring resonator with a radius of 10 $\mu$m in the add-drop configuration. The add-drop configuration is necessary to insert the microring sample in a loop.  The measured quality factor of the ring resonances is of several thousands (Q=2500$-$3000), with a free spectral range (FSR) of 7.5 nm. In Fig. \ref{fig:2} the transmission spectra of the add-drop ring resonator for both the through (a) and the drop (b) ports are shown along with the resonances chosen for the stimulated FWM experiment. The resolution of the spectra is 50 pm. In Fig. \ref{fig:2} (a), the resonances go down to below 5\% of transmission, meaning that the ring resonator is close to the critical coupling condition.

The source of gain is a Booster Optical Amplifier (Thorlabs BOA1004P), with a small signal gain of up to 30 dB.  The cavity is closed on the add and drop ports of the microring resonator. Input and output coupling with the silicon chip is obtained through the use of grating couplers. The add-drop resonator basically acts, inside the cavity, as a band-pass filter (BPF) for each resonance. In our experiment, a single resonance ($\lambda_p = 1555.87$ nm) was selected for lasing by restricting the cavity transmission using an external BPF at the amplifier output. This external filter also serves to reduce amplified spontaneous emission (ASE) from the BOA to a level lower than the generated FWM experiments. In order to obtain the BOA background noise suppression, we cascaded three BPFs for a total rejection of more than 150 dB in the idler generation band.
A second BPF, identical to the previous one, tuned to the pump resonance was used at the sample output to further improve rejection of frequencies other than the lasing mode before amplification. The laser power inside the cavity was monitored using a 99:1 beam splitter (BS) before the amplifier.


\begin{figure}
\includegraphics[width=\columnwidth]{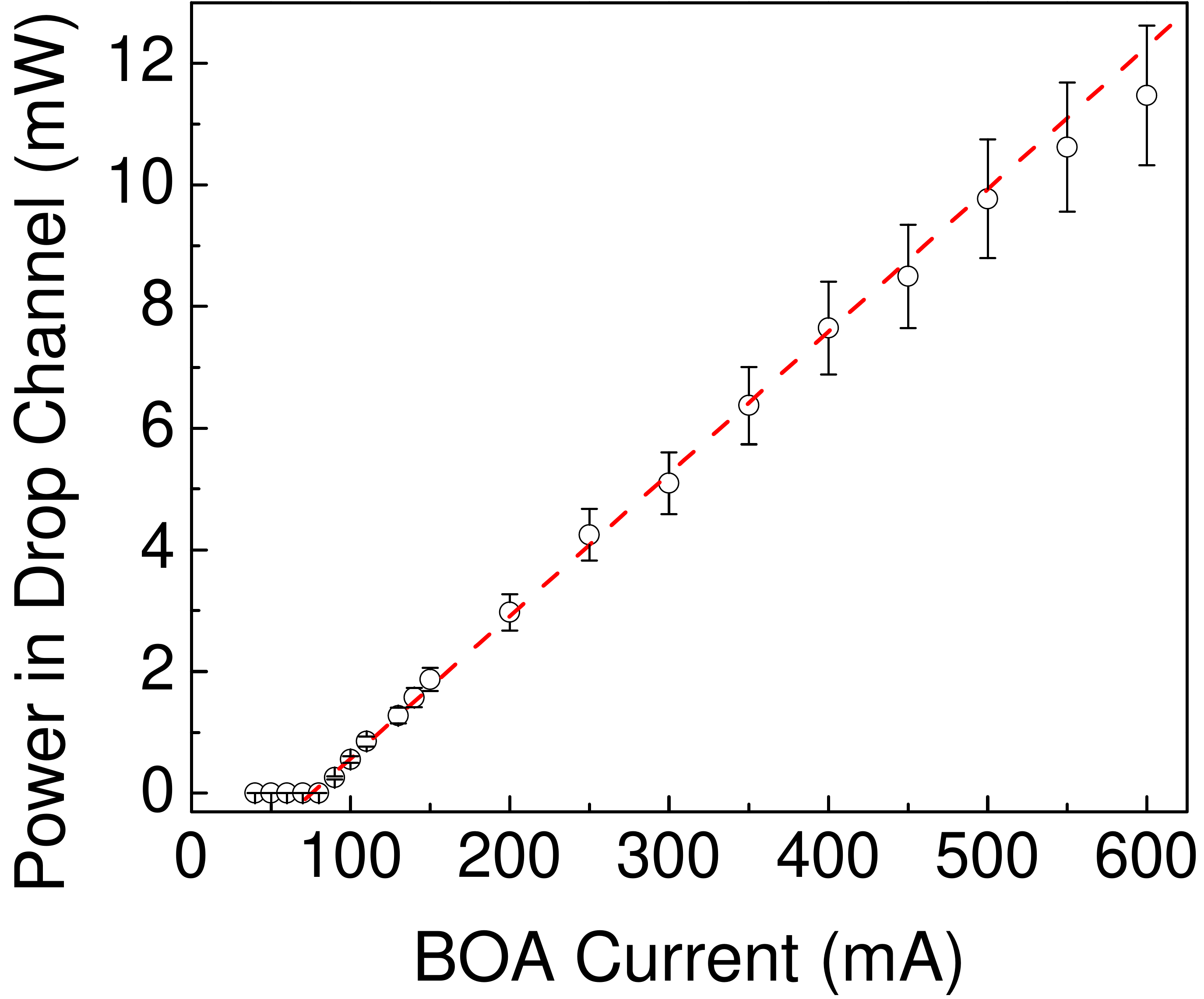}
\caption{Lasing curve of the cavity. The black circles are the experimental values, whereas the red line shows the linear fit corresponding to Eq. (\ref{eq:LasingCurve}).  }
\label{fig:3}
\end{figure}

The lasing curve of the cavity is shown in Fig. \ref{fig:3} as a function of the BOA current. On the y-axis the optical power estimated at the drop port of the ring resonator is reported.
It is calculated from the measured power at the 1\% port of the 99:1 BS and compensating for both BPF and coupling losses. A clear threshold behavior is observed around a current $i_{th}=90$ mA. The total losses for one loop, excluding the BOA, were directly measured in the cavity to be -18 dB and are distributed among the different components in the following way: the coupling losses are 3.6 dB for each grating coupler, the BPFs losses are estimated to be 3.5 dB for each filter, the 50:50 BS losses are 3 dB, whereas the losses due to the isolator and the 99:1 BS are 0.3 dB and 0.5 dB, respectively.
Indeed, the lasing threshold corresponds to a current where the BOA small signal gain is 20 dB, close to the inverse of the losses.

In Fig. \ref{fig:3}, it is also reported the linear fit of the characteristic curve of the cavity according to the following equation \cite{Rigrod:1965,Eckbreth:1975}:
\begin{equation}
\label{eq:LasingCurve}
P_{out} \left( I \right) = P_{sat} \left( kI-g_{th} \right) \frac{G_{th}}{G_{th}-1} T_{tot} \left( 1-T_{oc} \right),
\end{equation}
where $P_{out}$ is the optical power read at the output of the 1\% port of the 99:1 BS, I is the BOA current,  $P_{sat}$ is saturation power,  $G_{th}=e^{g_{th}}$ is the saturated gain of the amplifying medium, $g_0=kI$ with $g_0$ small signal gain,  $T_{tot}$ is the total transmission of the optical elements in the cavity (except for the 99:1 BS transmission) and  $T_{oc}$ is the transmission of the output coupler, intended as transmission internal to the ring resonator. The figure shows that the experimental points deviate from the linear fit for high BOA current. This might be due to the onset of the two-photon absorption process in the microring resonator when high power is stored inside it, but a more accurate study on this effect is needed.

\begin{figure}
\includegraphics[width=\columnwidth]{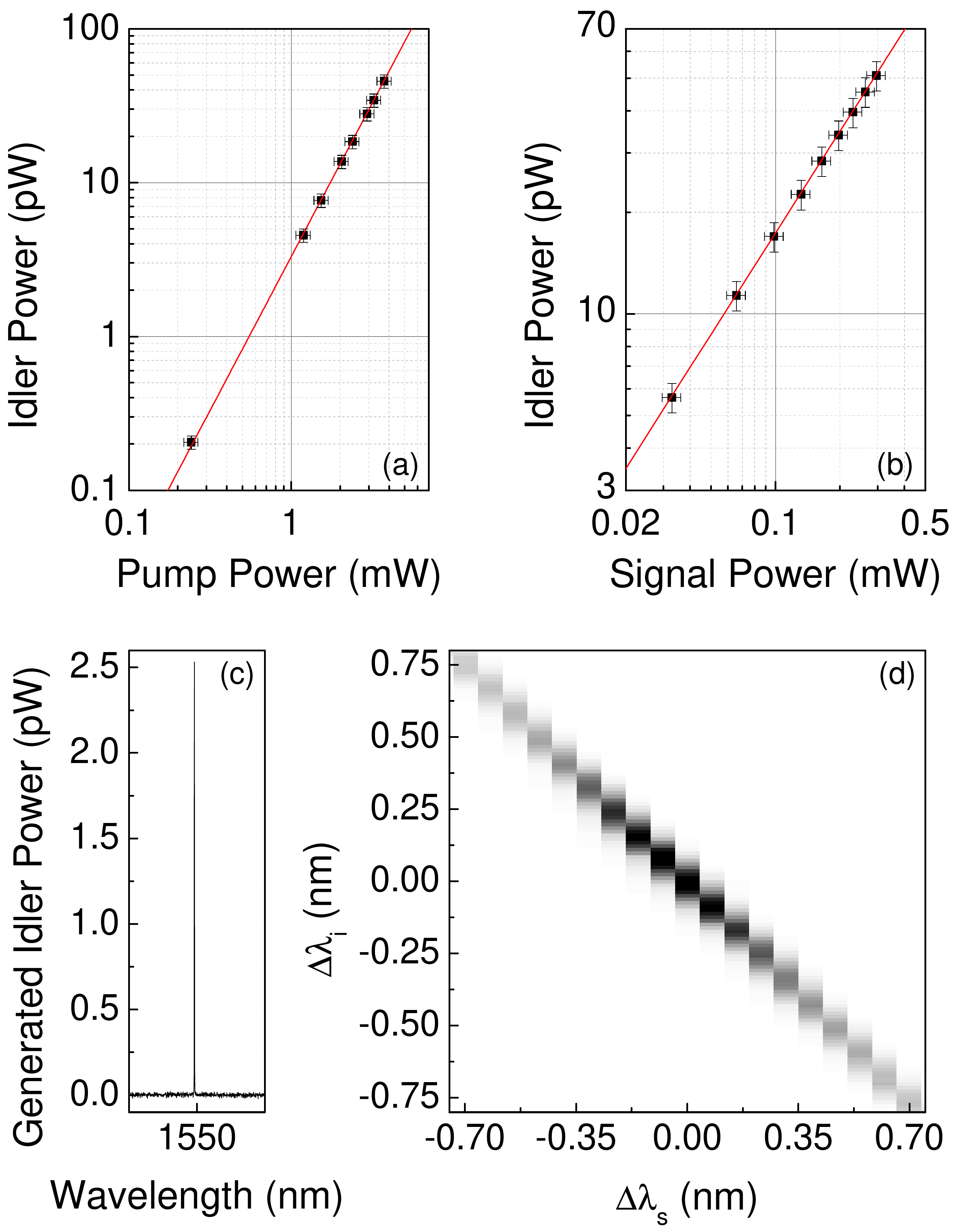}
\caption{ (a) and (b) show the stimulated FWM generation rates. The idler generation rate is proportional to the square of the lasing power (a) and grows linearly with the signal power (b) inside the ring resonator. In (a) the signal power coupled to the sample is 130 $\mu$W, whereas in (b) the BOA current is fixed at 250 mA. Red lines corresponding to the correct slope required by the stimulated FWM process are added to the graphs. (c) One idler spectrum taken for coupled pump and signal powers of 1.87 mW and 130 $\mu$W respectively. (d) Idler-signal correlation measurement as a function of the wavelengths corresponding to the idler and signal resonances. The measured intensity is closely peaked around the antidiagonal, showing clear energy-time correlations. The spectral resolution on the idler resonance is given by the receiving spectrometer and is 67 pm.}
\label{fig:4}
\end{figure}

Given the low quality factor of the present resonator, spontaneous FWM was too weak to be observed. We performed instead the classical FWM experiment resonantly exciting a resonance (signal) of the ring at a wavelength of 1563.45 nm thus producing a stimulated idler beam at a wavelength of 1548.39 nm. In Fig. \ref{fig:2} the resonances used for the FWM experiment are displayed (I, P and S stand for idler, pump and signal respectively). As shown in Fig. \ref{fig:1}, the external tunable CW infrared laser (Santec TSL-510) is coupled with the ring cavity using a 50:50 BS just before the sample and is completely suppressed by the BPF placed before the BOA so that it has no loop gain. Moreover, the signal laser is spectrally cleaned before being injected into the cavity by means of a tunable band-pass filter (Santec OTF-350) in order to remove spurious ASE photons. The idler photons are finally collected on a spectrometer equipped with a liquid-nitrogen cooled CCD camera.
The conversion efficiencies of the stimulated FWM process are shown in Fig. \ref{fig:4} (a) and (b). The idler generation rate is proportional to the square of the lasing power (a) and grows linearly with the signal power (b) inside the ring resonator, proving its parametric origin. In Fig. \ref{fig:4} (c) an example of FWM spectrum corresponding to a pump power coupled inside the ring of 1.87 mW and a coupled signal power of 130 $\mu$W is also shown.

The classical FWM intensity can be used to directly assess the energy correlations between the signal-idler photon pairs that would be emitted in the spontaneous parametric process \cite{Grassani:2016:JSD:scirep}. Indeed, the linewidth of the ring resonances is of the order of several tens of GHz and is much larger that the spacing of the modes of the laser loop cavity, which is several meters in length. This means that the pump laser could encompass many lasing modes and span the whole linewidth. However, the emission of time-entangled photons requires a pump linewidth smaller than that of signal and idler photons \cite{Helt2010} and this might not be the case in the used self-pumping geometry. On the other hand, gain narrowing is expected to occur within the relatively large bandwidth of the microring resonance, solving the problem of having a laser line narrower than the signal and idler resonances.

The idler-signal correlation curve can be directly assessed via a joint spectral density (JSD) measurement \cite{Liscidini:2013,Grassani:2016:JSD:scirep,Fang:Liscidini:2014:fastJSD,Eckstein:LasPhotRev:2014,Jizan:Helt:Eggleton:16:JSDwithphase,Silverstone2015}. To perform this measurement, we acquire FWM spectra on a CCD camera by varying the signal wavelength from 1560 nm to 1566 nm in steps of 10 pm. Spectral resolution on the idler resonance is given by the receiving spectrometer and is 67 pm. During the measurement, the BOA current is kept fix to 200 mA, whereas the optical power of the signal inside the ring resonator is 250 $\mu$W . The correlation measurements are shown in Fig. \ref{fig:4} (d) as a function of the signal and idler resonances' relative wavelengths. The measured intensity is closely peaked around the anti-diagonal, showing clear correlations between the signal and idler's energies. This proves that, at least with the current microring quality factor, signal-idler photon pairs emitted in the spontaneous process would be entangled. 

In summary, we have proved that FWM can be achieved in a silicon integrated microring resonator without the need of an external laser. These results are an important step towards the realization of a silicon-based source of entangled photons that does not require an external pump, making it closer to real world applications.
In order to have a potentially usable emission rate ($>$ MHz) for the spontaneous process, a ring resonator with a quality factor exceeding 10000 is needed. Indeed, silicon ring resonators with such a high quality factor have been already employed for studying a variety of effects, including optical bistability and FWM \cite{AlmeidaLipsonOL2004,Turner2008,Azzini2012OL,Engin2013,Grassani:Optica:2015}.
The present result is a strong indication that the emitted photons would be entangled, even if a definitive proof will consist in a Franson experiment \cite{Franson1991,Grassani:Optica:2015} performed under self-pumping conditions.

Another interesting possible output for our system would be to have a separable state, in order to produce single photons via heralding \cite{Davanco2012}. In order to achieve this, the laser linewidth should be comparable to the linewidth of the ring resonator’s mode \cite{Liscidini:2013}. This can be obtained by inserting a modulator inside the cavity to broaden the laser emission. In the case of a ring resonator with a quality factor of 40000, this corresponds to a linewidth of about 5 GHz, a frequency easily achievable with state-of-the-art fiber modulators. In such a scheme, one should achieve a state purity comparable to the limit for optical pumping \cite{Helt2010}.

Moreover, a recently very active research direction is the possibility to achieve emission of quantum states of light of more than two photons \cite{Reimer:Morandotti:Science:2016,Llewellyn_Thompson_CLEO_2018_multiphoton,1803.01641,LiscidiniGHZ-PRAppl2017}. Our self-pumping system can reach enough CW power inside the ring resonator to achieve the multiphoton emission regime. Indeed, a power of 5 mW inside a ring resonator of Q$\sim$10000 is sufficient to have an emission rate of multiple pairs exceeding $10^5$ Hz inside the ring \cite{Azzini2012OE,Takesue2010}. Furthermore, multiple photon emission can be obtained by building the laser cavity so to achieve pulsed laser emission. Pulsing in fiber loop cavities has been already accomplished using several different geometries \cite{RUNGE2014657,Yang:04_Opt_expr_Sub_pico_pulsed}, which could be directly ported to our approach \cite{Roztocki:2017}.

Finally, we would like to point out that the scheme presented in our work could be directly applied to an all-pass ring resonator by considering the backscattering at the input port. 
In fact, the backscattered light would behave as the light coming out the drop port of an add-drop resonator and it could be of the order of -5 dB, depending on the coupling factor \cite{Morichetti:2010}. This, in turn, would make closing an all-pass microring in a fiber cavity possible.
\bigskip

The Integrated Photonic Technologies Center (INPHOTEC) of Scuola Superiore Sant'Anna in Pisa is acknowledged for PIC manufacturing. Moreover, we thank the University of Pavia Blue Sky Research under Project No. BSR1732907 for support. This research was also supported by the Italian Ministry of Education, University and Research (MIUR): ``Dipartimenti di Eccellenza Program (2018-2022)", Department of Physics, University of Pavia.

\end{document}